# Hanle effect missing in a prototypical organic spintronic device

Alberto Riminucci [a),*], Mirko Prezioso [a)], Chiara Pernechele [b)], Patrizio Graziosi [a),c)], Ilaria Bergenti [a)], Raimondo Cecchini [a)], Marco Calbucci [a)], Massimo Solzi [b)], V. Alek Dediu [a)]

a)Consiglio Nazionale delle Ricerche - Istituto per lo Studio dei Materiali Nanostrutturati (CNR-ISMN), Bologna, Italy

b)Università di Parma, Dipartimento di Fisica e Scienze della Terra, Parma, Italy

c)Instituto de Tecnología de Materiales, Universidad Politécnica de Valencia, Valencia, Spain

*email: a.riminucci@bo.ismn.cnr.it

## Abstract

We investigate spin precession (Hanle effect) in the prototypical organic spintronic giant magnetoresistance (GMR) device $La_{0.7}Sr_{0.3}MnO_3$(LSMO)/tris(8-hydroxyquinoline)(Alq3)/AlOx/Co. The Hanle effect is not observed in measurements taken by sweeping a magnetic field at different angles from the plane of the device. As possible explanations we discuss the tilting out of plane of the magnetization of the electrodes, exceptionally high mobility or hot spots. Our results call for a greater understanding of spin injection and transport in such devices.

# Introduction

The field of organic spintronics has been the subject of vigorous experimental and theoretical investigation since its inception[1-9] because of its great scientific appeal and applicative potential.[10-12] Although much progress has been made in explaining interfacial properties,[13-17] progress on the understanding of spin transport in the organic medium met greater difficulties.

Over a short time organic spintronics has put together an impressive record of different phenomena observed, such as organic magnetoresistance,[18] spin polarization inversion at the organic semiconductor/ferromagnetic interface,[1] organic tunneling anisotropic magnetoresistance[19] and magnetically modulated memristance,[10]

In particular organic giant magnetoresistance (GMR) devices,[8, 20-24] in which the thickness of the organic layer is too great for tunneling between the electrodes to occur, require spin injection and transport in the organic semiconductor.

All these phenomena involve the measurement of a magnetoresistance (MR), i.e. resistance as a function of an applied magnetic field, whether the device involved tunneling across an organic barrier or instead injection of carriers in the organic semiconductor.

MR by itself cannot be considered as proof of spin injection unless the Hanle effect is detected.[25] This effect is caused by the precession of the spin in the presence of a non collinear magnetic field and can be detected by the way it affects the MR. The Hanle effect was measured in spintronic devices with an inorganic semiconductor spin transporting channel[26-30] and established spin injection in inorganic semiconductors on a sure footing.

Despite the fact that the mechanisms for charge carrier transport in inorganic semiconductors are different from those of the organic ones, there are no fundamental reasons to exclude the observation of the Hanle effect in the latter.

In this article we report on the investigation of the Hanle effect on an organic GMR device (figure 1) that consists in a $La_{0.7}Sr_{0.3}MnO_3$(LSMO)/tris(8-hydroxyquinoline)(Alq3)/AlOx/Co stack, where Alq3 is the spin transporting medium in which spin precession can take place. These devices posses memristive[31] properties, that is, their resistance is dependent on the history of the applied current or voltage;[10, 32] in the same device, different resistive states posses different GMR. This system has been extensively studied and its properties are well known[8, 20, 21, 24, 33] and is therefore ideal to investigate the presence of the Hanle effect.

Alq3 is an electron transporting medium, due to the electrons' higher mobility compared to holes[34] and due to the energy alignment at the interface.[35, 36] In the remainder we will therefore assume that current is only carried by electrons.

Experimental evidence[37] demonstrate that when a non collinear magnetic field is applied, spin precession must occur in unpaired electrons in organic semiconductors, in particular in Alq3.[38] The applied magnetic field has to be at least as strong as those intrinsic to the material (e.g. the hyperfine field)[2] for the Hanle effect to unfold.

To be best of our knowledge this is the first work on the investigation of the Hanle effect in organic GMR devices.

## Results and discussion

The device, with an active area of 1×1 mm$^2$ and schematically depicted in figure 1 was fabricated by depositing the LSMO bottom electrode (20 nm thick) by channel spark ablation (CSA) in $O_2$ at $2\times10^{-2}$ mbar on a matching $SrTiO_3$ substrate. Alq3 (200 nm) was evaporated from an effusion cell with the substrate kept at room temperature in an ultra high vacuum chamber at a base pressure of $10^{-8}$ mbar. The film was amorphous with a roughness of about 1 nm rms. The AlOx barrier (2.5 nm thick) was deposited by CSA in an Ar atmosphere at $2\times10^{-2}$ mbar. Co (20 nm), the top ferromagnetic electrode, was evaporated using an electron gun at a base pressure of $10^{-8}$ mbar. It must be mentioned that the AlOx tunnel barrier is necessary for the top interface to have good morphological and chemical properties.[13-15] All measurements in the following were taken at 100 K and the MR was measured by applying a bias voltage of -100 mV to the LSMO electrode. Care was taken to exclude possible artifacts due to the MR of the electrodes.[39]

Depending on the relative orientation of the two magnetic electrodes, the device can be in a high resistance state $R_P$ (parallel magnetization of the electrodes) or in a low resistance state $R_{AP}$ (antiparallel magnetization). When the direction of magnetization of one of the electrodes flips, the MR switches abruptly between $R_P$ and $R_{AP}$. The GMR is quantified as the ratio between the difference in resistance of the two states and the resistance of the parallel state:

$$GMR = \frac{R_{AP} - R_P}{R_P}$$

The Hanle effect in spintronic devices can show itself in two ways, depending on how it affects the GMR of the device. When spin transport is incoherent, it is usually detected by measuring the depolarizing effect it has on a spin polarized current, which in turn causes a decrease of the GMR.[28] When instead transport is sufficiently coherent, each precession can be observed as an oscillation in the resistance as a function of the applied magnetic field.[40] In the following, since charge transport in organic semiconductors takes place by hopping[41] and is therefore highly incoherent, we will look for the former behavior.

In our experiments the magnetic field was applied at various angles $\theta$ from the plane of the device ($\vartheta=0$ means that the field is applied in plane, as described in figure 1). In this geometry the GMR is proportional to:

$$GMR \propto \int_0^{+\infty} \frac{1}{\sqrt{4\pi Dt}} \cdot e^{-\frac{(L-vt)^2}{4Dt}} \cdot [\sin^2\theta \cos(\omega_P t) + \cos^2\theta] dt$$

where $L$= 200 nm is the thickness of the organic layer, $D$ is the diffusion constant, $v$ is the drift velocity, and $\omega_P$ is the angular frequency.[42-45] In the case at hand, as we suppose decoherent transport, the integral with the $cos(\omega_P t)$ term vanishes and the GMR is proportional to $\cos^2\theta$. Measurements were carried out by sweeping the magnetic field between -0.3 T and 0.3 T while keeping $\theta$ constant. Figure 2 shows the results for $\theta = 90^0$. As $\cos^2(90^0) = 0$ no GMR should be observed in this transport regime. Figure 2 instead shows two distinct resistances for the electrodes in the parallel (high resistance) and antiparallel (low resistance) configuration with a GMR=-6.7%; the full sweep is reported in the inset as well as the MR at $\theta = 0^0$. These results show that the Hanle effect is not observed. In the full sweep at $\theta = 90^0$ the linear MR at higher field can be attributed to the tilting out of plane of the magnetization of the electrodes.

In order to confirm the absence of the Hanle effect in our measurements, we also measured the GMR as a function of $\theta$ in a different memristive state, with a GMR=-18%. The low resistance state was reached by applying a positive voltage pulse to the device. Figure 3 summarizes the results of such experiments, carried out with the field applied in plane ($\theta = 0^0$), at $\theta = 45^0$ and $\theta = 60^0$, at which angle the highest switching field exceeded the available range. In each curve the resistance has a jump every time the coercive field of one of the electrodes is reached. In the antiparallel state of each curve, the resistance increases as the magnitude of the field increases, but this behavior does not fit that expected from a Hanle signal. The most important point is that figures 3b) and 3d) show no dependence of the GMR on $\theta$ apart from the switching fields, a fact that is not related to the Hanle effect.

In these measurements the MR goes between $R_{AP}$ and $R_P$ in more than one step. This can be explained if we think of the device as a collection of two parallel spintronic sub-devices, as schematically drawn in figures 3a) for the leftward magnetic field sweep and 3c) for the rightward one. In these two figures the boxes represent two parallel spin valves and the arrows show the orientation of the magnetization in each sub-electrode. In figure 3a) each box corresponds to a resistance level in figure 3b), as indicated by the numbers between parentheses. The same holds for figures 3c) and 3d). The peculiar shape of the MR at $\theta = 60^0$ is explained by the fact that one of the switching fields exceeds the available range and is therefore pinned. The switching fields show a $1/\cos\theta$ behavior which is due to the magnetization reversal process of the electrodes.[46]

One obvious reason for the absence of signs of the Hanle effect could be that the magnetization of the electrodes is aligned to the applied magnetic field. In this case the spin of the electrons would also be aligned to the applied magnetic field and no precession could take place. In order to rule out this possibility, we measured the magnetization of a Co on Si/Alq3 (50 nm)/AlOx (2.5 nm) and of a LSMO film by SQUID magnetometry. Results are reported in figure 4 for LSMO and in figure 5 for Co. At $\theta = 90^0$ and 20 mT, which is the upper limit of the magnetic field in figure 2, the out-of-plane magnetization of LSMO and Co is negligible and can be discounted as the explanation of the observed behavior.

Another possible explanation would be that we are in the coherent transport regime, but the spins do not have the time to precess while they cross the organic layer due to their high velocity. At a field of 20 mT the time for a full precession, considering a g factor equal to that of a free electron (which is sufficiently accurate for our purposes, see for example Ref.47), is ~1.8 ns. With a 10% accuracy, using $cos(\omega_P t) = 90\%, \omega_P = 3.52 \times 10^9 \, rad/s$, the electrons would need to take less than ~0.13 ns to cross the 200 nm thick organic layer, in order for the spin precession to go undetected. With an applied bias of -100 mV, this requires a mobility of ~30 cm² V⁻¹ s⁻¹. At 100 K this corresponds to a diffusion constant of 0.27 cm²/s and a diffusion time of 1.5 ns. The calculated mobility is more than one order of magnitude greater than the highest reported for Alq3 (~1 cm² V⁻¹ s⁻¹).[48] If we consider a more realistic figure for mobility of electrons in Alq3 (~10⁻⁶ cm² V⁻¹ s⁻¹),[49] the coherent transport picture is untenable even in the case that the thickness of the organic layer were very small in some areas(but still sufficiently thick to prevent direct tunneling across Alq3, i.e. thicker than a few monolayers).[50] The mobility required would still need to be about 5 orders of magnitude greater than the latter figure we reported. For extreme narrowing of the organic layer, tunneling could occur across the organic layer.[24] In this case the Hanle effect would not be present. On the other hand hot spots, in which LSMO contacts directly the AlOx/Co top layers, would give a positive GMR[51] and can be ruled out since this is in contrast to the negative GMR we observed.

## Conclusions

In this work we investigated the Hanle effect, one of the main outstanding issues in organic spintronics. We have investigated it by measuring the GMR of a prototypical organic spintronic device at different angles between the device's plane and the magnetic field and we found no sign of its presence. Although we have no definitive explanation for this finding, an exceptionally high mobility (30 cm$^2$ V$^{-1}$ s$^{-1}$) would be sufficient to justify the present data. All together, these results strongly suggests that the current understanding of transport in organic GMR devices is not sufficiently developed to explain the absence of the observation of spin precession and supports the framework of transport occurring via high mobility, high conductivity channels.

## Acknowledgments

We would like to acknowledge Luis Hueso, Georg Schmidt and Ian Appelbaum for stimulating this research and Federico Bona and Mauro Paoletti for their invaluable technical support. This work was supported by the European project "HINTS" NMP3-SL-2011-263104.

1. C. Barraud, P. Seneor, R. Mattana, S. Fusil, K. Bouzehouane, C. Deranlot, P. Graziosi, L. Hueso, I. Bergenti, V. Dediu, F. Petroff and A. Fert, Nature Physics **6** (8), 615-620 (2010).
2. P. A. Bobbert, W. Wagemans, F. W. A. van Oost, B. Koopmans and M. Wohlgenannt, Physical Review Letters **102** (15), 156604 (2009).
3. M. Cinchetti, K. Heimer, J. P. Wustenberg, O. Andreyev, M. Bauer, S. Lach, C. Ziegler, Y. L. Gao and M. Aeschlimann, Nat. Mater. **8** (2), 115-119 (2009).
4. V. Dediu, M. Murgia, F. C. Matacotta, C. Taliani and S. Barbanera, Solid State Communications **122** (3-4), 181-184 (2002).
5. A. J. Drew, J. Hoppler, L. Schulz, F. L. Pratt, P. Desai, P. Shakya, T. Kreouzis, W. P. Gillin, A. Suter, N. A. Morley, V. K. Malik, A. Dubroka, K. W. Kim, H. Bouyanfif, F. Bourqui, C. Bernhard, R. Scheuermann, G. J. Nieuwenhuys, T. Prokscha and E. Morenzoni, Nat. Mater. **8** (2), 109-114 (2009).
6. L. E. Hueso, I. Bergenti, A. Riminucci, Y. Q. Zhan and V. Dediu, Advanced Materials **19** (18), 2639-2642 (2007).
7. P. P. Ruden and D. L. Smith, Journal of Applied Physics **95** (9), 4898-4904 (2004).
8. Z. H. Xiong, D. Wu, Z. V. Vardeny and J. Shi, Nature **427** (6977), 821-824 (2004).
9. S. Sanvito and V. A. Dediu, Nat Nano **7** (11), 696-697 (2012).
10. M. Prezioso, A. Riminucci, P. Graziosi, I. Bergenti, R. Rakshit, R. Cecchini, A. Vianelli, F. Borgatti, N. Haag, M. Willis, A. J. Drew, W. P. Gillin and V. A. Dediu, Advanced Materials, DOI: 10.1002/adma.201202031 (2012).
11. W. J. Baker, K. Ambal, D. P. Waters, R. Baarda, H. Morishita, K. van Schooten, D. R. McCamey, J. M. Lupton and C. Boehme, Nat Commun **3**, 898 (2012).
12. T. Miyamachi, M. Gruber, V. Davesne, M. Bowen, S. Boukari, L. Joly, F. Scheurer, G. Rogez, T. K. Yamada, P. Ohresser, E. Beaurepaire and W. Wulfhekel, Nat Commun **3**, 938 (2012).
13. A. A. Sidorenko, C. Pernechele, P. Lupo, M. Ghidini, M. Solzi, R. De Renzi, I. Bergenti, P. Graziosi, V. Dediu, L. Hueso and A. T. Hindmarch, Applied Physics Letters **97** (16), 162509 (2010).
14. Y. Q. Zhan, X. J. Liu, E. Carlegrim, F. H. Li, I. Bergenti, P. Graziosi, V. Dediu and M. Fahlman, Applied Physics Letters **94** (5), 053301 (2009).
15. F. Borgatti, I. Bergenti, F. Bona, V. Dediu, A. Fondacaro, S. Huotari, G. Monaco, D. A. MacLaren, J. N. Chapman and G. Panaccione, Applied Physics Letters **96** (4), 043306 (2010).
16. T. Methfessel, S. Steil, N. Baadji, N. Grossmann, K. Koffler, S. Sanvito, M. Aeschlimann, M. Cinchetti and H. J. Elmers, Physical Review B **84** (22), 224403 (2011).
17. Y. Q. Zhan and M. Fahlman, J. Polym. Sci. Pt. B-Polym. Phys. **50** (21), 1453-1462 (2012).
18. T. L. Francis, O. Mermer, G. Veeraraghavan and M. Wohlgenannt, New Journal of Physics **6** (185), 1-8 (2004).
19. M. Gruenewald, M. Wahler, F. Schumann, M. Michelfeit, C. Gould, R. Schmidt, F. Wuerthner, G. Schmidt and L. W. Molenkamp, Physical Review B **84** (12), 125208 (2011).
20. V. Dediu, L. E. Hueso, I. Bergenti, A. Riminucci, F. Borgatti, P. Graziosi, C. Newby, F. Casoli, M. P. De Jong, C. Taliani and Y. Zhan, Physical Review B **78** (11), 115203 (2008).
21. D. Sun, L. Yin, C. Sun, H. Guo, Z. Gai, X. G. Zhang, T. Z. Ward, Z. Cheng and J. Shen, Physical Review Letters **104** (23), 236602 (2010).
22. M. Palosse, M. Fisichella, E. Bedel-Pereira, I. Seguy, C. Villeneuve, B. Warot-Fonrose and J. F. Bobo, Journal of Applied Physics **109** (7) (2011).
23. S. Majumdar, R. Laiho, P. Laukkanen, I. J. Vayrynen, H. S. Majumdar and R. Osterbacka, Applied Physics Letters **89** (12), 122114 (2006).
24. W. Xu, G. J. Szulczewski, P. LeClair, I. Navarrete, R. Schad, G. X. Miao, H. Guo and A. Gupta, Applied Physics Letters **90** (7), 3 (2007).
25. F. G. Monzon, H. X. Tang and M. L. Roukes, Physical Review Letters **84** (21), 5022-5022 (2000).
26. N. Tombros, C. Jozsa, M. Popinciuc, H. T. Jonkman and B. J. van Wees, Nature **448** (7153), 571-574 (2007).
27. I. Appelbaum, B. Huang and D. J. Monsma, Nature **447** (7142), 295-298 (2007).
28. M. Johnson and R. H. Silsbee, Physical Review Letters **55** (17), 1790-1793 (1985).
29. J. M. Kikkawa and D. D. Awschalom, Physical Review Letters **80** (19), 4313-4316 (1998).

30. X. Lou, C. Adelmann, S. A. Crooker, E. S. Garlid, J. Zhang, K. S. M. Reddy, S. D. Flexner, C. J. Palmstrom and P. A. Crowell, Nature Physics **3** (3), 197-202 (2007).
31. L. Chua, Applied Physics A-Materials Science & Processing **102** (4), 765-783 (2011).
32. M. Prezioso, A. Riminucci, I. Bergenti, P. Graziosi, D. Brunel and V. A. Dediu, Advanced Materials **23** (11), 1371-1375 (2011).
33. A. Riminucci, I. Bergenti, L. E. Hueso, M. Murgia, C. Taliani, Y. Zhan and F. Casoli, arXiv:0701603 (2007).
34. H. H. Fong and S. K. So, Journal of Applied Physics **100** (9), 094502 (2006).
35. Y. Q. Zhan, I. Bergenti, L. E. Hueso and V. Dediu, Physical Review B **76** (4), 045406 (2007).
36. Y. Q. Zhan, M. P. de Jong, F. H. Li, V. Dediu, M. Fahlman and W. R. Salaneck, Physical Review B **78** (4), 045208 (2008).
37. J. M. Lupton, D. R. McCamey and C. Boehme, ChemPhysChem **11** (14), 3040-3058 (2010).
38. M. N. Grecu, A. Mirea, C. Ghica, M. Colle and M. Schwoerer, J. Phys.-Condes. Matter **17** (39), 6271-6283 (2005).
39. A. Riminucci, M. Prezioso, P. Graziosi and C. Newby, Applied Physics Letters **96** (11), 112505 (2010).
40. B. Huang, D. J. Monsma and I. Appelbaum, Physical Review Letters **99** (17), 177209 (2007).
41. V. Coropceanu, J. Cornil, D. A. da Silva, Y. Olivier, R. Silbey and J. L. Bredas, Chem. Rev. **107** (4), 926-952 (2007).
42. F. J. Jedema, M. S. Nijboer, A. T. Filip and B. J. van Wees, Physical Review B **67** (8), 085319 (2003).
43. M. Johnson and R. H. Silsbee, Physical Review B **37** (10), 5312-5325 (1988).
44. M. Johnson and R. H. Silsbee, Physical Review B **37** (10), 5326-5335 (1988).
45. J. Li, B. Huang and I. Appelbaum, Applied Physics Letters **92** (14), 142507 (2008).
46. J. M. D. Coey, *Magnetism and Magnetic Materials*. (Cambridge University Press, Cambridge, 2011).
47. Z. G. Yu, Physical Review B **85** (11), 115201 (2012).
48. A. J. Drew, F. L. Pratt, J. Hoppler, L. Schulz, V. Malik-Kumar, N. A. Morley, P. Desai, P. Shakya, T. Kreouzis, W. P. Gillin, K. W. Kim, A. Dubroka and R. Scheuermann, Physical Review Letters **100** (11), 116601 (2008).
49. B. J. Chen, W. Y. Lai, Z. Q. Gao, C. S. Lee, S. T. Lee and W. A. Gambling, Applied Physics Letters **75** (25), 4010-4012 (1999).
50. G. Szulczewski, H. Tokuc, K. Oguz and J. M. D. Coey, Applied Physics Letters **95** (20) (2009).
51. J. M. De Teresa, A. Barthelemy, A. Fert, J. P. Contour, F. Montaigne and P. Seneor, Science **286** (5439), 507-509 (1999).

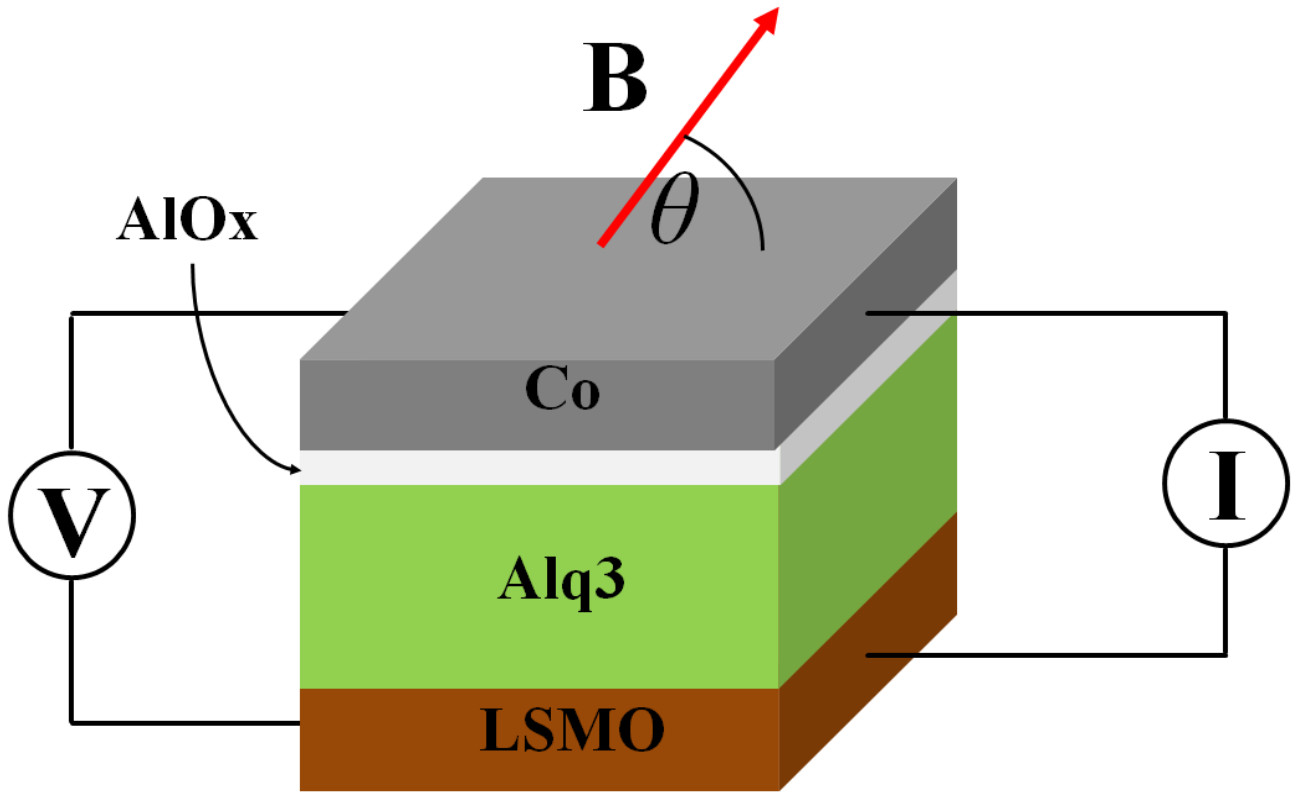

Figure 1: Schematic drawing of the organic GMR device. From top to bottom: Co ferromagnetic top electrode (20 nm), AlOx tunnel barrier (2.5 nm), Alq3 spin transporting organic semiconductor layer (200 nm), $La_{0.7}Sr_{0.3}MnO_3$ bottom ferromagnetic electrode (20 nm). The active area is 1×1 mm². A current I is driven through the device and a voltage V is measured. B is the applied magnetic field, at angle $\theta$ with the plane of the device.

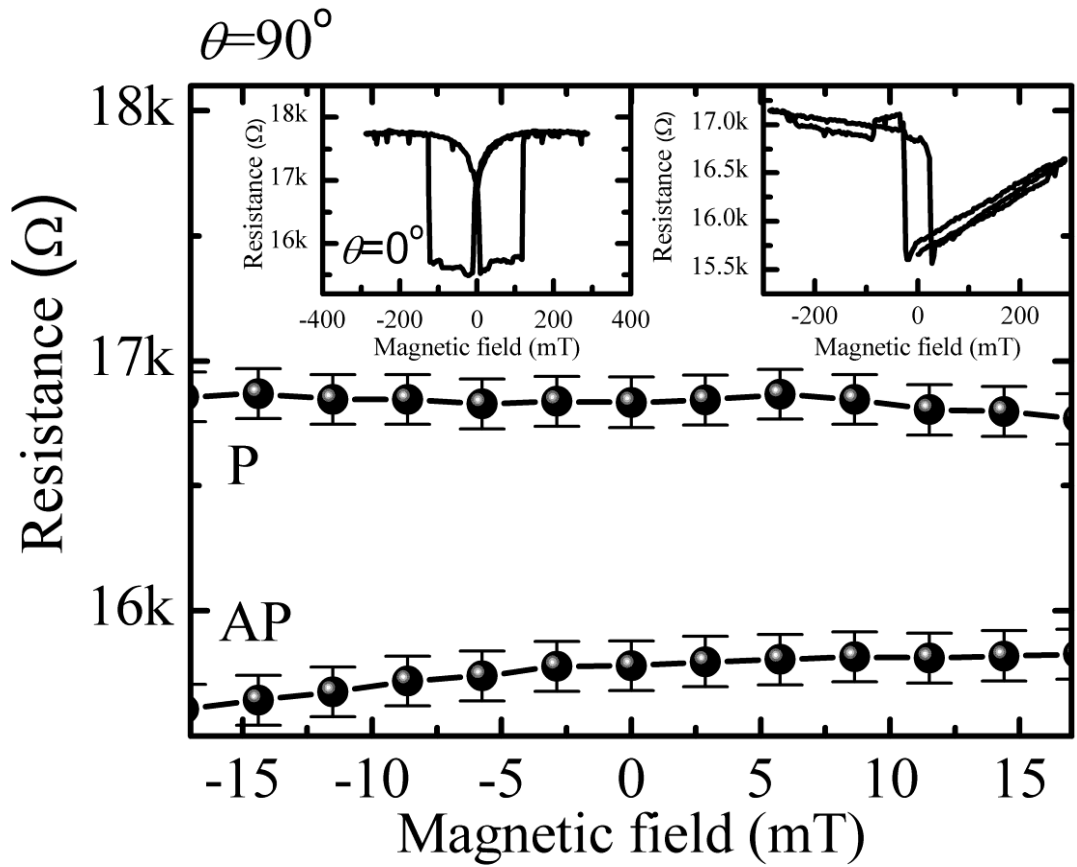

Figure 2: Resistance in the parallel (P) and antiparallel (AP) state of the device, with the magnetic field applied at angle $\theta=90^0$ from the plane of the device. The top left inset shows the MR when the field is applied in the plane of the device, while the inset to the top right shows the complete MR for $\theta=90^0$. The linear behavior at high field is due to the tilting out of plane of the magnetization of the electrodes.

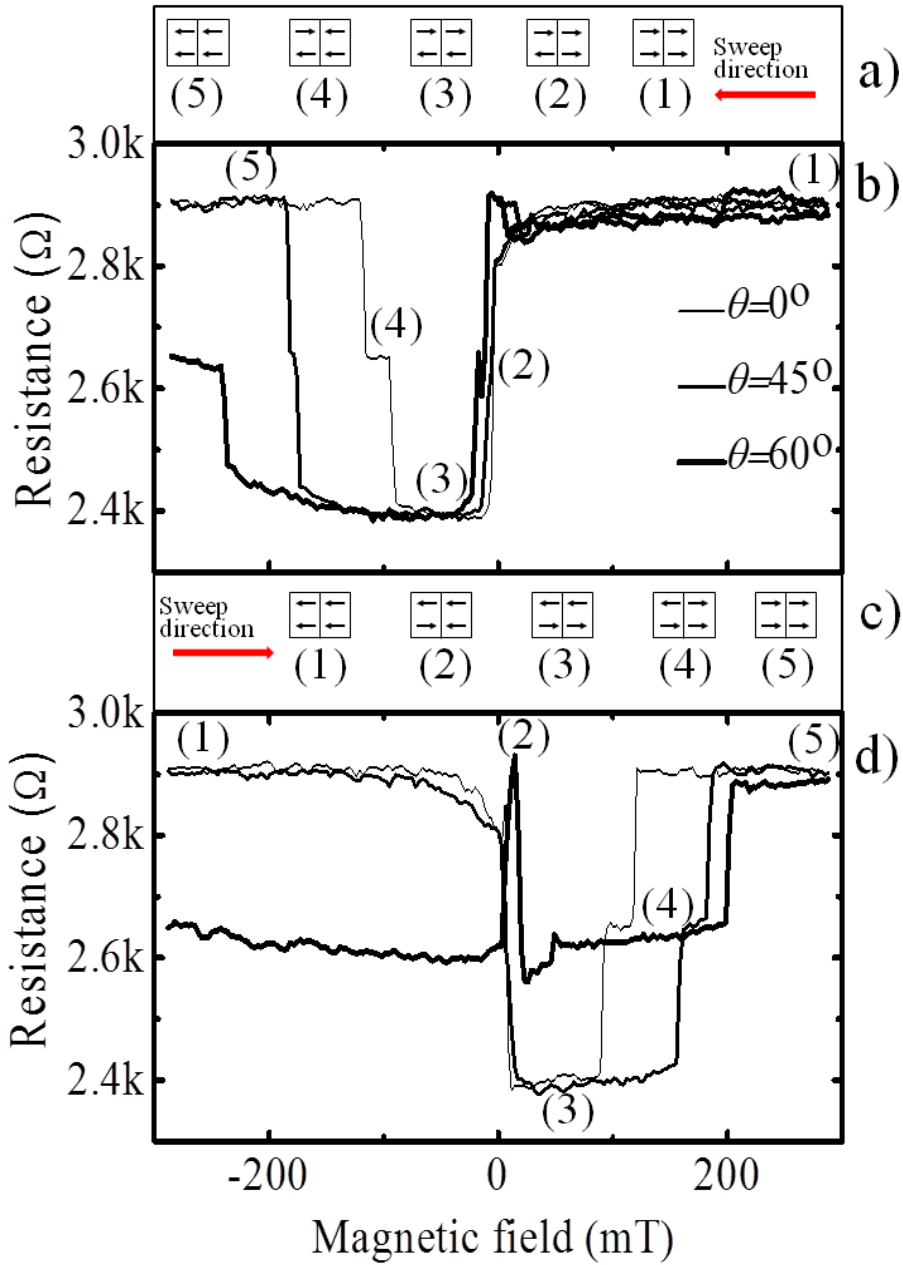

Figure 3: MR at $\theta=0$, $\theta=45^0$ and $\theta=60^0$ in a different resistive state from that in figure 2. In a) and c) the boxes are a schematic representation of the device as the combination of two parallel sub-devices. In each box, the arrows indicate the orientations of the magnetization in each sub-electrode. Each sub-electrode has a different coercive field. b) Leftward magnetic field sweep. The numbers between the parentheses indicate the resistance level, and correspond to the magnetization configuration indicated by the same number in a). c) Rightward magnetic field sweep. The numbers between parenthesis have the same meaning as above. The unusual look of the MR for $\theta=60^0$ is due to the fact that one of the switching fields exceeds the available range.

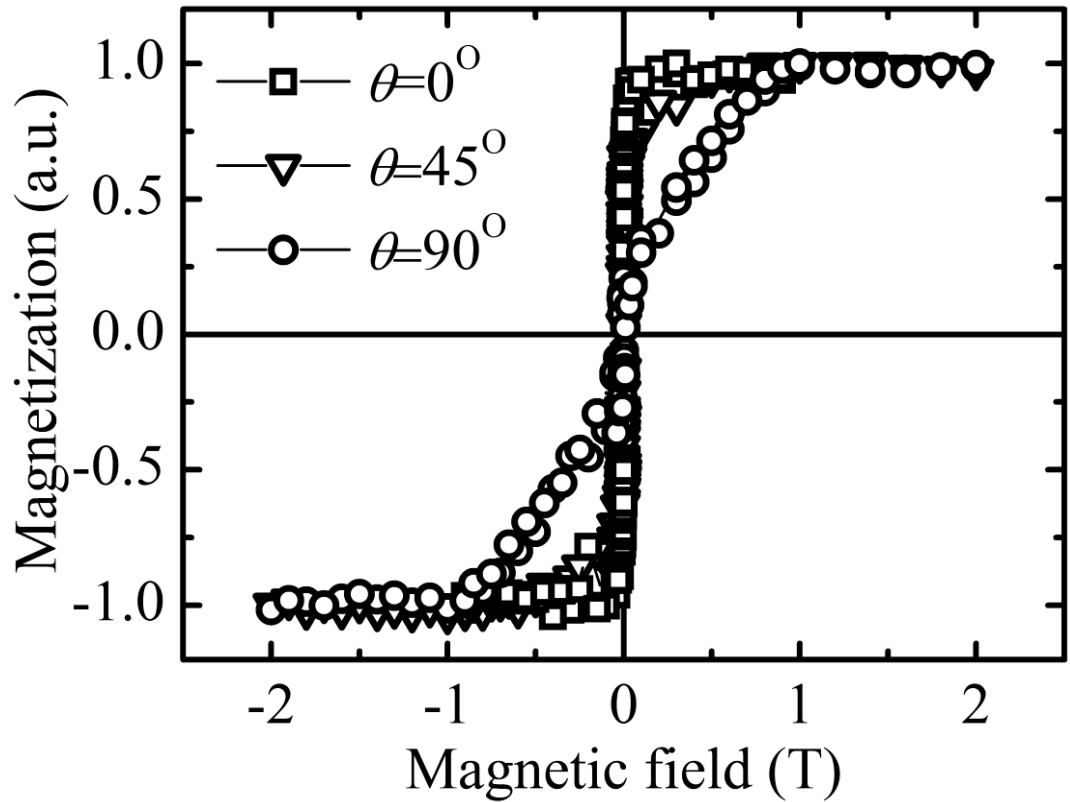

Figure 4: Hysteresis loops of a LSMO 20 nm thick film obtained by SQUID magnetometry at 100 K. The loop with square symbols corresponds to the magnetic field being applied in the plane of the film, the one with triangles to the field at $\theta=45^0$ with the plane, and the circles to the field at $\theta=90^0$. At $\theta=90^0$ and 20 mT the magnetization of the film is tilted out of plane by about 6%.

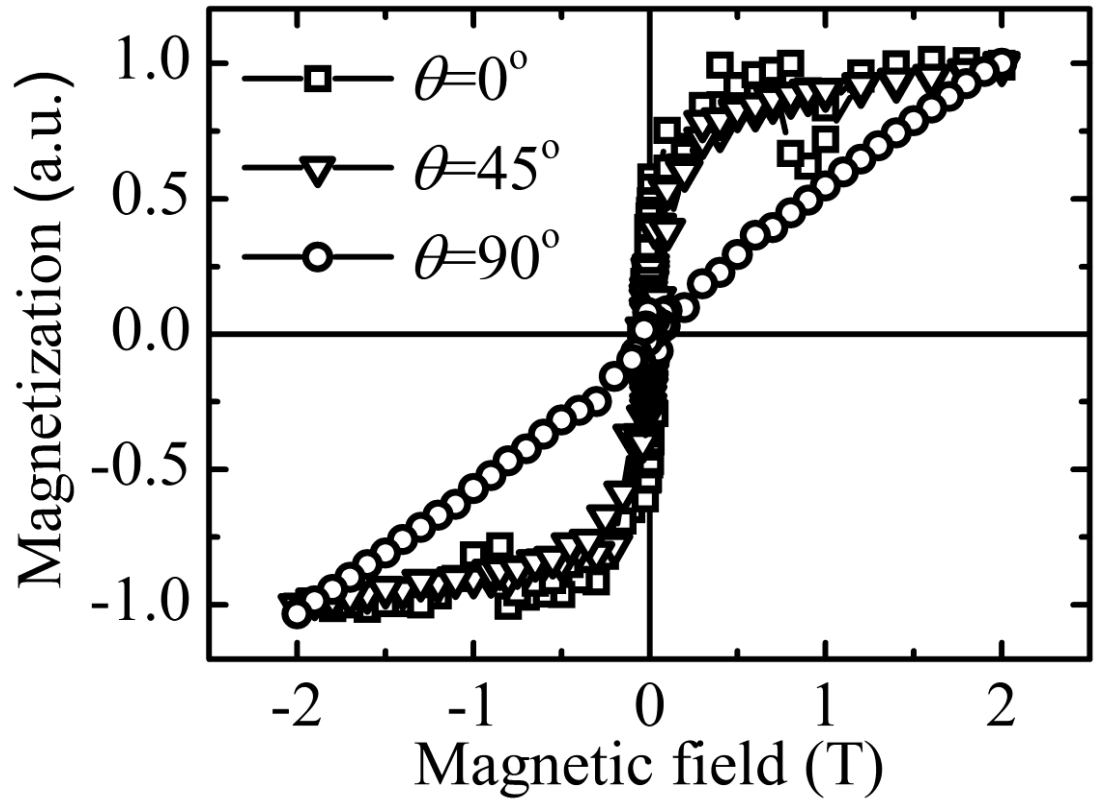

Figure 5: Hysteresis loops for a 20 nm thick Co film grown on a AlOx(2.5 nm)/Alq3 (50 nm)/Si, measured in a SQUID magnetometer at 100 K. The magnetic field was applied at $\theta=0$ (squares), $\theta=45^0$ (triangles) and $\theta=90^0$ (circles) from the plane of the device. At $\theta=90^0$ and 20 mT the magnetization is tilted out of plane by 1%.